\documentclass[conference]{IEEEtran}
\IEEEoverridecommandlockouts
\usepackage{cite}
\usepackage{amsmath,amssymb,amsfonts}
\usepackage{graphicx}
\usepackage{textcomp}
\usepackage[linesnumbered,lined,boxed,commentsnumbered, ruled]{algorithm2e}
\usepackage{epstopdf}
\def\BibTeX{{\rm B\kern-.05em{\sc i\kern-.025em b}\kern-.08em
    T\kern-.1667em\lower.7ex\hbox{E}\kern-.125emX}}
\usepackage[caption=false,font=footnotesize]{subfig}
\begin{document}

\title{Fully Decentralized Massive MIMO Detection Based on Recursive Methods}
\author{
	\IEEEauthorblockN{Jes\'{u}s Rodr\'{i}guez S\'{a}nchez, Fredrik Rusek, Muris Sarajli\'{c}, Ove Edfors and Liang Liu}\\
	\IEEEauthorblockA{Department of Electrical and Information Technology, Lund University, Sweden}
	\IEEEauthorblockA{\{jesus.rodriguez, fredrik.rusek, muris.sarajlic, ove.edfors, liang.liu\}@eit.lth.se}
}
\maketitle

\begin{abstract}
Algorithms for Massive MIMO uplink detection typically rely on a centralized approach, by which baseband data from all antennas modules are routed to a central node in order to be processed. In case of Massive MIMO, where hundreds or thousands of antennas are expected in the base-station, this architecture leads to a bottleneck, with critical limitations in terms of interconnection bandwidth requirements. This paper presents a fully decentralized architecture and algorithms for Massive MIMO uplink based on recursive methods, which do not require a central node for the detection process. Through a recursive approach and very low complexity operations, the proposed algorithms provide a sequence of estimates that converge asymptotically to the zero-forcing solution, without the need of specific hardware for matrix inversion. The proposed solution achieves significantly lower interconnection data-rate than other architectures, enabling future scalability.
\end{abstract}


\section{Introduction}
\label{section:intro}
\begin{figure*}\centering
	
	\subfloat[Centralized architecture]{
		\includegraphics[width=0.4\textwidth]{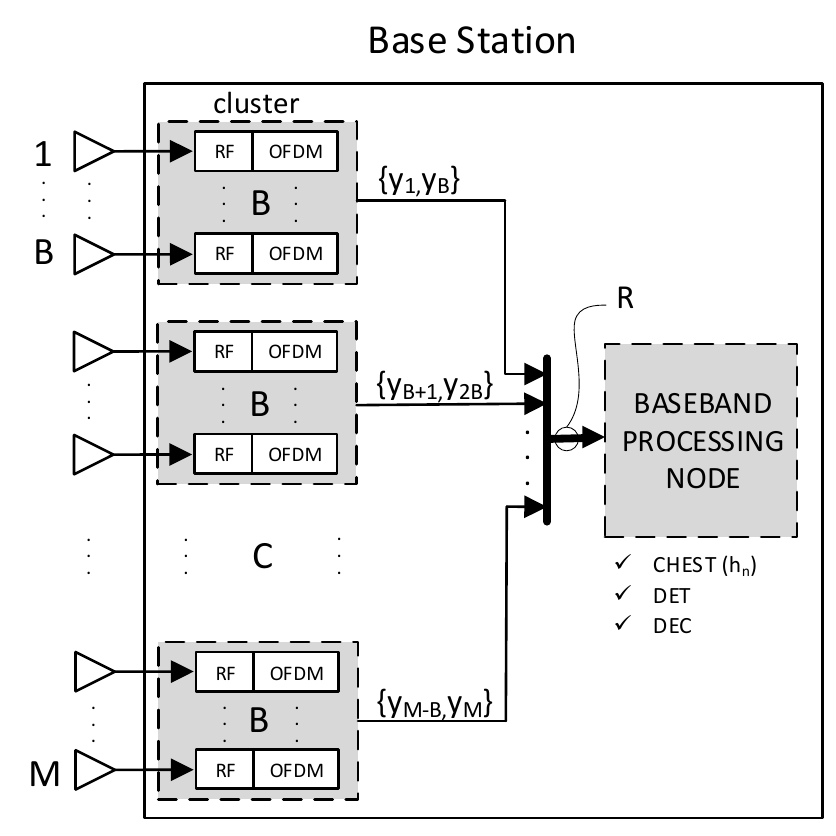}
		\label{fig:BS_centralized}
	}
	\subfloat[Decentralized architecture]{
		\includegraphics[width=0.4\textwidth]{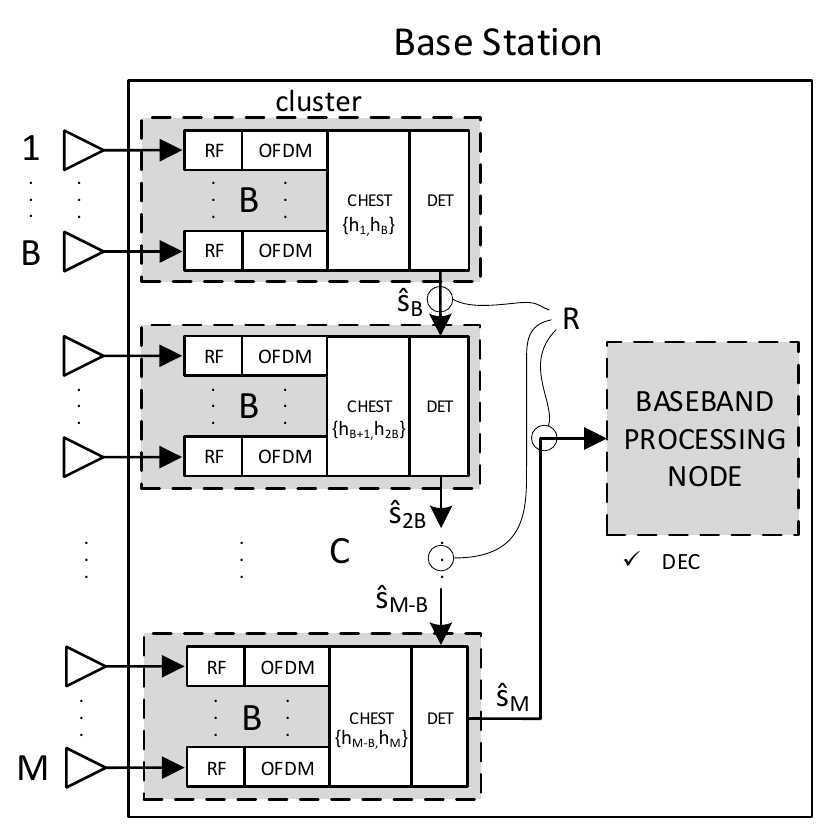}
		\label{fig:BS_decentralized}
	}
	
	\caption{Comparison between base station receiver chain in centralized and fully decentralized architectures for Massive MIMO uplink. Antenna array with M elements is divided into C clusters, each containing B antennas. (a): Centralized architecture. Clusters contain RF amplifiers and frequency down-conversion (RF) elements, analog-to-digital converters (ADC) and OFDM receivers. Each cluster has one link to transfer baseband samples to a central baseband processing node, where the rest of processing tasks are done. (b): Fully decentralized architecture for detection. Clusters performs RF, ADC, OFDM, channel estimation (CHEST) and detection (DET) locally. Decoding (DEC) is centralized. Clusters are connected to each other by uni-directional links. Only one cluster has a direct connection with central node. Proposed algorithms are executed in DET blocks in parallel mode. The points where the interconnection data-rate is estimated are marked by circles and the value is denoted by $\mathrm{R}$.}
	\label{fig:comparison}
\end{figure*}

Massive multi-user (MU) multiple-input multiple-output (MIMO) is one of the most promising technologies in the wireless area \cite{marzetta}. High spectral efficiency and improved link reliability are among the key features of this technology, making it a key enabler to exploit spatial diversity far beyond traditional MIMO systems by employing a large scale antenna array with hundreds or thousands of elements. This allows for unprecedented spatial resolution and high spectral efficiency, while providing simultaneous service to several users within the same time-frequency resource.

Despite all advantages of Massive MIMO, there are challenges from an implementation point of view. Uplink detection algorithms like zero-forcing (ZF) typically rely on a centralized architecture, shown in Figure  \ref{fig:BS_centralized}, where baseband samples and channel state information (CSI) are collected in the central node for further matrix inversion and detection. Dedicated links are needed between antenna modules and central node to carry this data. This approach, that is perfectly valid for a relatively low number of antennas, shows critical limitations when the array size increases, with interconnection bandwidth quickly becoming a bottleneck in the system. 

Previous work has been done proposing different architectures for Massive MIMO base-stations \cite{argos,larsson,puglielli,lumami,cavallaro}. All of them conclude by pointing to the interconnection bandwidth as the main implementation bottleneck and a limiting factor for array scalability.
Most of them recommend moving to a decentralized approach where uplink detection and downlink precoding can be performed locally in processing nodes close to the antennas. However, to achieve that, CSI still needs to be collected in a central node, where matrix inversion is done and the result distributed back to all modules \cite{argos,larsson,lumami}. A further step has been made in \cite{cavallaro}, where CSI is obtained and used only locally (not shared) for precoding and detection. This architecture relies on a central node only for processing partial results. This dependency on a central node limits the scalability of this solution as will be shown in section \ref{section:analysis}.

In this paper we propose a fully decentralized architecture and recursive algorithms for Massive MIMO uplink detection. Antennas in the array are grouped into clusters. Apart from antennas, clusters contain RF, Analog-to-Digital Converters (ADC), OFDM receiver, channel estimation and detection blocks. The decentralized topology is based on the direct connection of clusters forming a daisy-chain structure as shown in Figure \ref{fig:BS_decentralized}. The proposed algorithms are pipelined so that they run in a distributed way at the cluster nodes, providing a sequence of estimates that converge asymptotically to the zero-forcing solution. We will make use of the following algorithms: Recursive Least Square (RLS), Stochastic Gradient Descent (SGD) and Averaged Stochastic Gradient Descent (ASGD), which are detailed in section \ref{section:algorithms}.

Decentralized architectures overcome bottlenecks by finding a more equal distribution of the system requirements among the processing nodes of the system. Apart from this, data localization is a key characteristic of decentralized architectures. This architecture allows data to be consumed as close as possible to where it is generated, minimizing the amount to transfer, and therefore saving bandwidth and energy. Following this idea, processing nodes need to be located near the antenna. Further, they perform tasks such as channel estimation and detection locally. Local CSI is estimated and stored locally in each, without any need to share it with any other nodes in the system. 

The remainder of the paper is organized as follows. The system model for MIMO uplink is presented in section \ref{section:model}. In section \ref{section:algorithms} we introduce the proposed algorithms. In \ref{section:analysis} we analyze the performance of these algorithms, present the advantages of the daisy-chain topology, and analyze interconnection data-rates. Finally, section \ref{section:conclusions} presents the conclusions of this publication.

Notation: In this paper, lowercase, bold lowercase and upper bold face
letters stand for scalar, column vector and matrix, respectively. The
operations $(.)^T$, $(.)^*$ and $(.)^H$ denote transpose, conjugate and conjugate transpose respectively.
The vector $\mathbf{s}$ in the $n$th iteration is $\mathbf{s}_n$. Computational complexity is measured in terms of the number of complex-valued multiplications.\\

%%%%%%%%%%%%%%%%%%%%%%%%%%%%%%%%%%%%%%%%%%%%%%%%%%%%%%%%%%%%%%%%%%
\section{System model and detection algorithms}
\label{section:model}
In this section we present the system model for MIMO uplink and introduce the ZF equalizer.

We consider a scenario with $K$ single-antenna users transmitting to an antenna array with $M$ elements. The input-output relation for uplink is
\begin{equation}
\mathbf{y} = \mathbf{H}\mathbf{s} + \mathbf{v},
\label{eq:ul_model}
\end{equation}
where $\mathbf{y}$ is the $M \times 1$ received vector, $\mathbf{s}$ is the transmitted user data vector ($K \times 1$), $\mathbf{H}=[\mathbf{h}_1 \; \mathbf{h}_2 \, \cdots \, \mathbf{h}_M]^{{T}}$ is the channel matrix ($M \times K$) and $\mathbf{v}$ samples of noise ($M \times 1$). Under the Massive MIMO assumption, $M \gg K$.

Assuming time-frequency-based channel access, a Resource Element (RE) represents a slot in the time-frequency grid. Within each RE, the channel model follows \eqref{eq:ul_model}.

A least-squares (LS) estimate of  $\mathbf{s}$ is obtained as
\begin{equation}
\hat{\mathbf{s}}_{\mathrm{ZF}} = (\mathbf{H}^H \mathbf{H})^{-1}\mathbf{H}^H \mathbf{y}.
\label{eq:LS_sol}
\end{equation}
This method, commonly referred to as ZF, requires a central architecture as in Figure \ref{fig:BS_centralized} because the complete matrix $\mathbf{H}$ needs to be collected in the central node before the Gramian matrix ($\mathbf{H}^H\mathbf{H}$) and its inverse can be computed. Decentralized architectures, such as the one shown in Figure \ref{fig:BS_decentralized}, require other type of algorithms.
%%%%%%%%%%%%%%%%%%%%%%%%%%%%%%%%%%%%%%%%%%%%%%%%%%%%%%%%%%%%%%%%%%

\section{Proposed Algorithms}
\label{section:algorithms}
In this section we propose three algorithms for MIMO decentralized uplink detection.

Depending on the situation some algorithms are more appropriate than others. If full knowledge of matrix $\mathbf{H}$ and $\mathbf{y}$ is available at a single node, direct methods such as ZF can be applied \eqref{eq:LS_sol}. However, there are situations when the cost of collecting all knowledge at a single node is too high. For those cases, a different approach has to be used.

The goal of the proposed algorithms for uplink detection is the estimation of the transmitted user data vector, $\mathbf{s}$ in \eqref{eq:ul_model}, based on  knowledge of $\mathbf{H}$ and $\mathbf{y}$ that is distributed among nodes. These algorithms provide a sequence of estimates, which converge to $\hat{\mathbf{s}}_{ZF}$ as more knowledge of $\mathbf{H}$ and $\mathbf{y}$ is obtained. Estimation is done in a sequential manner, by which the estimate is passed from one antenna to the next one, being updated every time based on the previous estimate ($\hat{\mathbf{s}}_{n-1}$), local CSI ($\mathbf{h}_n$) and antenna observation ($y_n$). This can be summarized as $\hat{\mathbf{s}}_{n} = f(\hat{\mathbf{s}}_{n-1}, \mathbf{h}_n, y_n)$, which can be seen as a recursive form. This approach is in accordance with the data localization principle, which is a key characteristic of decentralized systems. In the Massive MIMO case, data is consumed close to where it is generated, namely at the antennas. This makes it possible that neither $\mathbf{h}_n$ nor $y_n$ are shared, since only the estimate is.

These algorithms are flexible enough to work in clusters of antennas (see Figure \ref{fig:BS_decentralized}), whose size can vary from 1 up to $M$, the last case being equivalent to a centralized system. 

The first recursive algorithm to be presented is the Recursive Least Square (RLS) method, which is a recursive form of \eqref{eq:LS_sol}. Uplink detection can be also seen as a regression parameter estimation -  a problem well studied in the area of stochastic approximation methods.
Stochastic Gradient Descent (SGD) and its averaged version (ASGD)  fall within this group, and are based on a Gradient Descent algorithm in which the gradient is partially known.

In Section \ref{section:RLS} we present  RLS  applied to MIMO uplink detection, which provides approximate ZF performance at the expense of a preprocessing stage. Afterwards, we present the SGD algorithm and its enhanced version, the Averaged SGD (ASGD), which increases  robustness of SGD while achieving performance close to ZF for very large arrays.

Before we describe the algorithms we clarify the role played by the variable $B$, i.e., the number of antennas per cluster. Our algorithms are in fact independent of the value of $B$, therefore we present them with notation tailored to the choice $B=1$. However, $B>1$ is still of importance from an implementation point of view since each cluster may be implemented with a single processing unit. Thus, with $M=100$ antennas, the choice $B=1$ requires 100 processing units, while $B=10$ merely requires 10 such units. Nevertheless, performance of our algorithms remains the same. $B$ therefore takes a trade-off role: The larger the $B$, the less number of processing units, but meanwhile, the architecture becomes more centralized.

\subsection{Recursive Least-Squares (RLS)}
\label{section:RLS}
RLS is the recursive version of the LS algorithm. It can be shown \cite{book:Ljung} that the ZF/LS estimate, i.e., the l.h.s. of \eqref{eq:LS_sol}, can be approximated by the RLS as $\hat{\mathbf{s}}_{\mathrm{ZF}} \approx \hat{\mathbf{s}}_{M}$ where $\hat{\mathbf{s}}_n$ is recursively found as follows
\begin{equation}
\begin{aligned}
\varepsilon_n &= y_{n} - \mathbf{h}_{n}^T \hat{\mathbf{s}}_{n-1} \\
\mathbf{\Gamma}_{n} &= \mathbf{\Gamma}_{n-1} - \frac{\mathbf{\Gamma}_{n-1} \mathbf{h}_{n}^{*} \mathbf{h}_{n}^T \mathbf{\Gamma}_{n-1}}{ 1 + \mathbf{h}_{n}^T \mathbf{\Gamma}_{n-1} \mathbf{h}_{n}^{*} },\\
\hat{\mathbf{s}}_{n} &= \hat{\mathbf{s}}_{n-1} + \mathbf{\Gamma}_{n} \mathbf{h}_{n}^* \varepsilon_{n}. \\
\end{aligned}
\label{eq:RLS_sn}
\end{equation}
The quality of the approximation depends on the initial value of $\hat{\mathbf{s}}_0$. Nevertheless, for a randomly chosen $\hat{\mathbf{s}}_0$, the impact of $\hat{\mathbf{s}}_0$ quickly fades out over the index $n$ and it can be shown that $s_{M} \to \hat{\mathbf{s}}_{\mathrm{ZF}}$ as $M\to \infty$ with probability one. In \eqref{eq:RLS_sn}, $\hat{\mathbf{s}}_n$ is a $K \times 1$ vector and is the output of cluster $n$, $y_n$ is the observation at the \textit{n}th antenna, $\varepsilon_n$ is the prediction error and $\mathbf{\Gamma}_{n}$ is a $K \times K$ matrix. As a side comment, we remark that $\hat{\mathbf{s}}_n$ is an approximate LS solution up to the \textit{n}th antenna element.

In view of Figure \ref{fig:BS_decentralized}, increasing the iteration number in \eqref{eq:RLS_sn} from $n$ to $n+1$ corresponds to passing on information from cluster $n$ to cluster $n+1$. Each cluster receives an estimate of the transmitted data vector from previous cluster, $\hat{\mathbf{s}}_{n-1}$, and compute a new estimate $\hat{\mathbf{s}}_{n}$ based on local CSI, $\mathbf{h}_{n}$, and a local observation, $y_{n}$.

Under the block fading channel model, multiple Resource Elements (RE) in a certain region of the time-frequency grid experience identical channels. We name this region Coherence Block (CB), and following this model it is possible to re-use same CSI for all REs in the same CB.

Straightforward implementation of \eqref{eq:RLS_sn} at every RE is not efficient. In fact, a hefty share of the operations associated to \eqref{eq:RLS_sn} can be reused within the CB, namely those associated to computation of $\mathbf{\Gamma}_n$. Defining $\mathbf{\Gamma}_0=\mathbf{I}_K$ and
\begin{equation}
\begin{aligned}
	\mathbf{z}_{n} &= \mathbf{\Gamma}_{n-1} \mathbf{h}_{n}^* \nonumber \\
	\alpha_{n} &= \frac{1}{1 + \mathbf{h}_{n}^T \mathbf{z}_n}\nonumber \\
	\mathbf{\Gamma}_{n} &= \mathbf{\Gamma}_{n-1} - \alpha_n^{} \mathbf{z}_n^{} \mathbf{z}_n^H, \quad n=1,2,\ldots,M
\end{aligned}
\end{equation}
we see that at each RE it  suffices to compute
\begin{equation}
\begin{aligned}
	\varepsilon_n &= y_{n} - \mathbf{h}_{n}^T \hat{\mathbf{s}}_{n-1} \nonumber \\
	\hat{\mathbf{s}}_{n} &= \hat{\mathbf{s}}_{n-1} + \alpha_n \mathbf{z}_n \varepsilon_n, \quad n=1,2,\ldots,M 
\end{aligned}
\end{equation}
in order to execute \eqref{eq:RLS_sn}. It is easily verifiable that the complexity of preprocessing is $\mathcal{O}(K^2)$, whilst the complexity is $\mathcal{O}(K)$ at every RE.

\subsection{Stochastic Gradient Descent (SGD)}
The setup in SGD \cite{book:Kushner} is that one intends to solve the unconstrained LS problem
\begin{equation} \label{eq:SGD_R} \min_{\mathbf{s}} \|\mathbf{y}-\mathbf{H}\mathbf{s}\|^2\end{equation}
via a gradient descent (GD) approach. The gradient of \eqref{eq:SGD_R} equals $\nabla_{\mathbf{s}}=\mathbf{H}^{{H}}\mathbf{H}\mathbf{s}-\mathbf{H}^{{H}}\mathbf{y}.$ An immediate consequence is that GD is only feasible in a centralized approach.

SGD is an approximate version that can be operated in a decentralized architecture. It does so by computing, at each cluster, as much as possible of $\nabla_{\mathbf{s}}$ with the information available at the cluster. Then the cluster updates the estimate $\hat{\mathbf{s}}$ using a scaled version of the "local" gradient and passes the updated estimate on to the next cluster.
 
The above described procedure can, formally, be stated as
\begin{equation}
\begin{aligned}
\varepsilon_n &= y_{n} - \mathbf{h}_{n}^T \hat{\mathbf{s}}_{n-1} \\
\hat{\mathbf{s}}_{n} &= \hat{\mathbf{s}}_{n-1} + \mu_n \mathbf{h}_{n}^* \varepsilon_n,\\
\end{aligned}
\label{eq:SGD_sn}
\end{equation}
where $\{\mu_n\}$ is a sequence of scalar step-sizes.

\subsection{Averaged Stochastic Gradient Descent (ASGD)}
Selection of optimum values $\mu_n$ in SGD is not trivial. Even though we take $\mu_n = \mu$ for simplification, the optimum value will depend on $M$, $K$ and channel properties, where the latter may be unknown in many cases. An inappropriate selection of $\mu$ can lead to severe performance degradation depending on the scenario. Averaging a SGD sequence provides an asymptotically optimal convergence rate provided that the noise $\mathbf{v}$ is Gaussian \cite{polyak}, which increases robustness to the step-size selection. In the ASGD algorithm there are three sequences defined as follows
\begin{equation}
\begin{aligned}
\varepsilon_n &= y_{n} - \mathbf{h}_{n}^T \hat{\mathbf{x}}_{n-1} \\
\hat{\mathbf{x}}_{n} &= \hat{\mathbf{x}}_{n-1} + \mu_n \mathbf{h}_{n}^* \varepsilon_n\\
\hat{\mathbf{s}}_{n} &= 
	\begin{cases}
		\hat{\mathbf{x}}_n & \text{if } n < n_0\\
		\frac{1}{n-n_0+1} \sum_{k=n_0}^{n} \hat{\mathbf{x}}_k & \text{if } n \geq n_0,\\
	\end{cases}
\end{aligned}
\label{eq:ASGD}
\end{equation}
where $\hat{\mathbf{x}}_{n}$ takes the role of the SGD output $\hat{\mathbf{s}}_n$ in \eqref{eq:SGD_sn}. The ASGD output $\hat{\mathbf{s}}_{n}$ thereby becomes an averaged SGD sequence, where $n_0$ determines the onset of the averaging procedure.

The averaged sequence can be written more conveniently as
\begin{equation}
\hat{\mathbf{s}}_{n} = 
\begin{cases}
\hat{\mathbf{x}}_n & \text{if } n < n_0\\
\hat{\mathbf{s}}_{n-1} + \frac{1}{n'} \left(\hat{\mathbf{x}}_{n} - \hat{\mathbf{s}}_{n-1}\right) & \text{if } n \geq n_0,\\
\end{cases}
\label{eq:ASGD2}
\end{equation}
where $n'=n-n_0+1$. As will be seen in our numerical results, the ASGD grossly relaxes the need for careful selection of $\mu$.

\section{Analysis}
\label{section:analysis}
In this section we analyze the proposed solution. First, the performance of the presented algorithms will be shown and compared with each other. Second, a few strong points of the daisy-chain topology are given. Finally, an analysis of interconnection bandwidth is presented, followed by a comparison for four different configurations.

\subsection{Detection Performance}
In this section, we present performance results for all algorithms. Reported metrics are Mean-Square-Error (MSE) and Bit-Error-Rate (BER) in block faded Rayleigh channels.

We report MSE, measured between $\hat{\mathbf{s}}$ and $\mathbf{s}$, as a function of the number of iterations (antenna index). The reported signal-to-noise ratio (SNR) is the average receive power at any base station antenna, divided by the noise variance. 

\begin{figure}[h]
	\centering
	\includegraphics[width=0.35\textwidth]{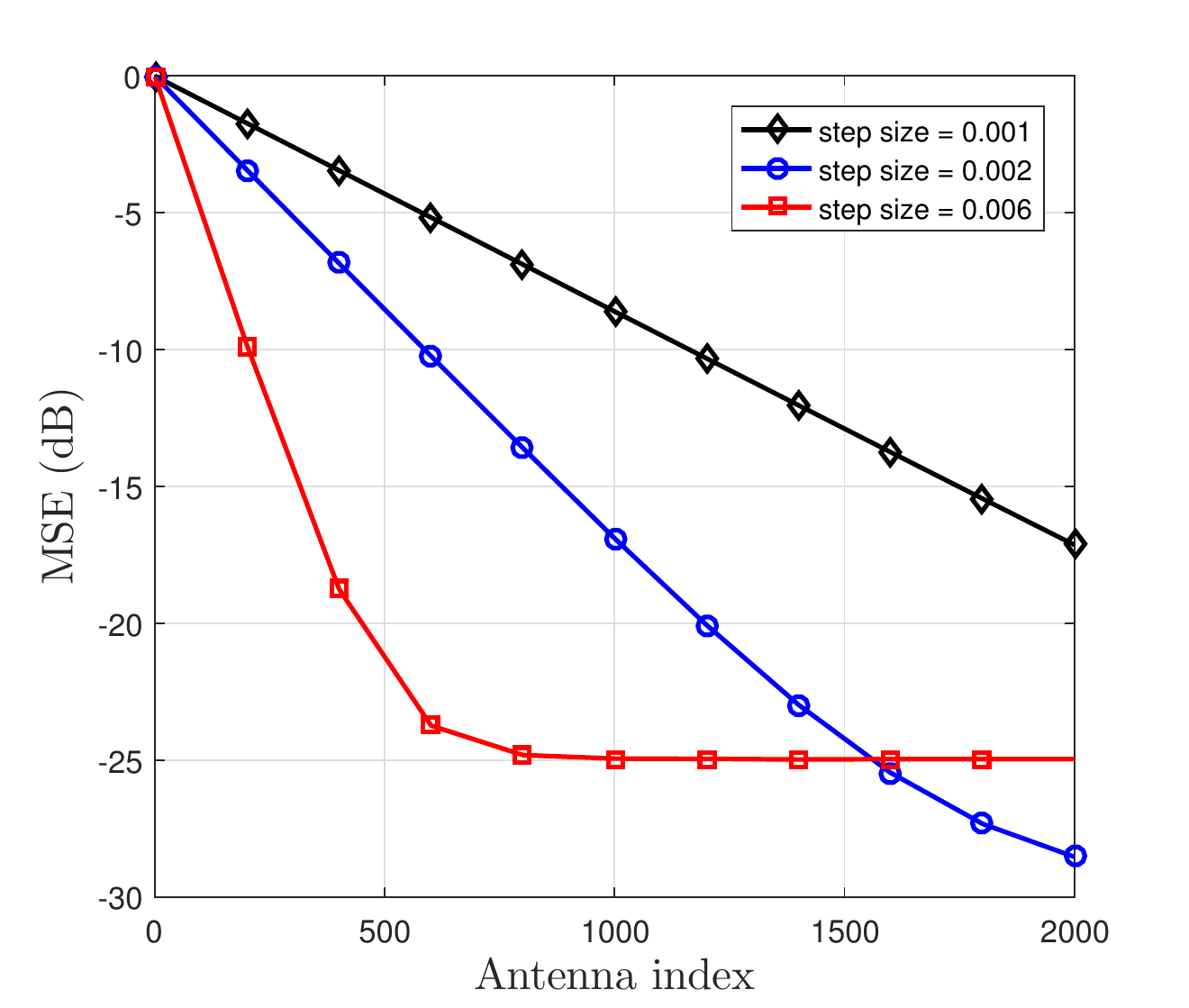}
		\vspace*{-3mm}
	\caption{MSE vs antenna index for three different step-size values in SGD.}
	\label{fig:MSE_vs_antenna_SGD_mu}
\end{figure}

MSE results for SGD are shown in Figure \ref{fig:MSE_vs_antenna_SGD_mu} for three different step-size values. As can be observed, step-size plays a critical role in the convergence speed of the algorithm. High step-size values provide faster convergence but high steady-state MSE, and low values may not even enter into the steady-state within the array. Given a certain $M$ and $K$, it is possible to find an optimum step-size which provides the lowest MSE.

\begin{figure*}[t!]
	\centering
	\subfloat{
		\includegraphics[width=0.4\textwidth]{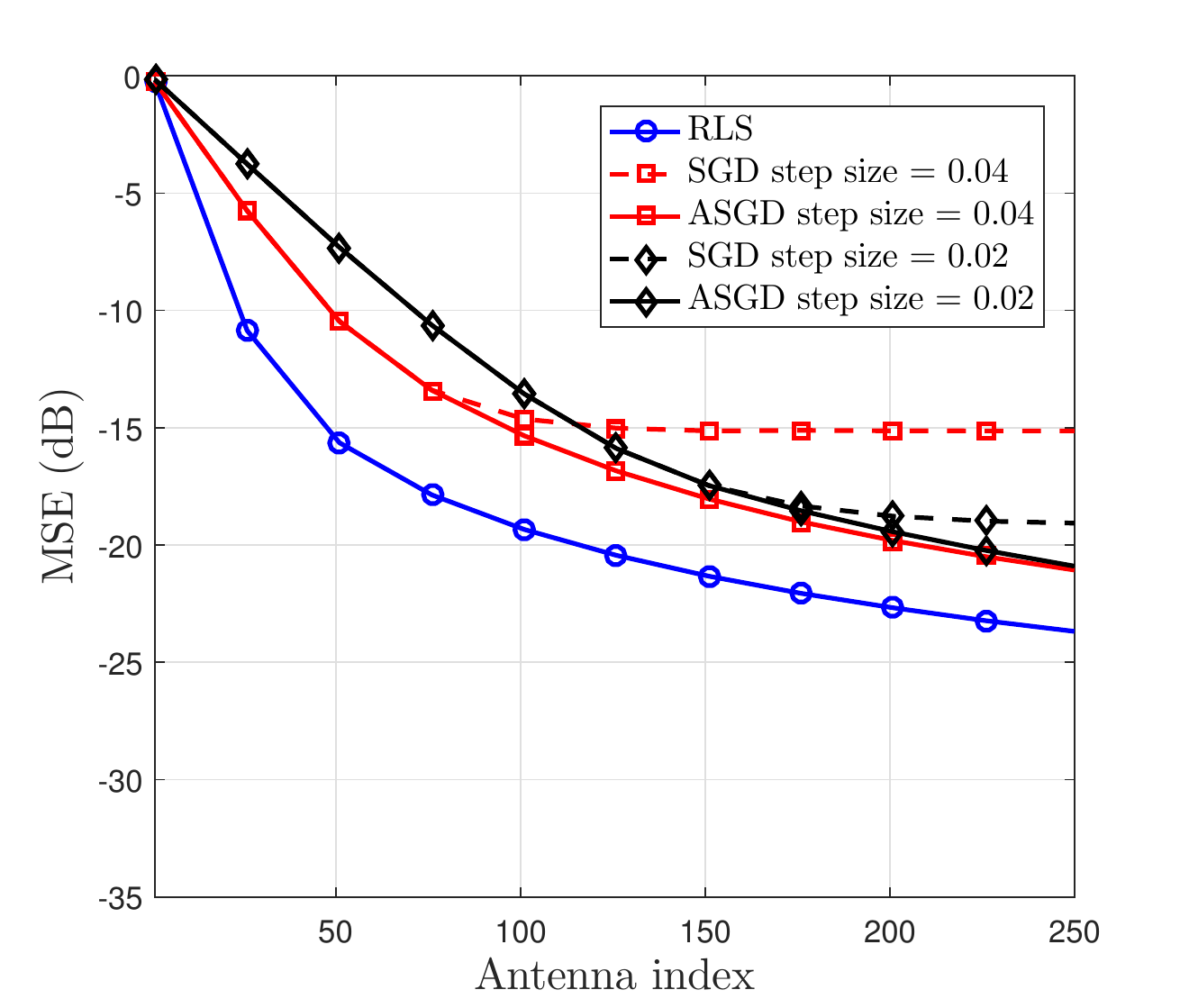}
		\label{fig:MSE_vs_antenna_RLS_SGD_ASGD_M256}
	}
	\subfloat{
		\includegraphics[width=0.4\textwidth]{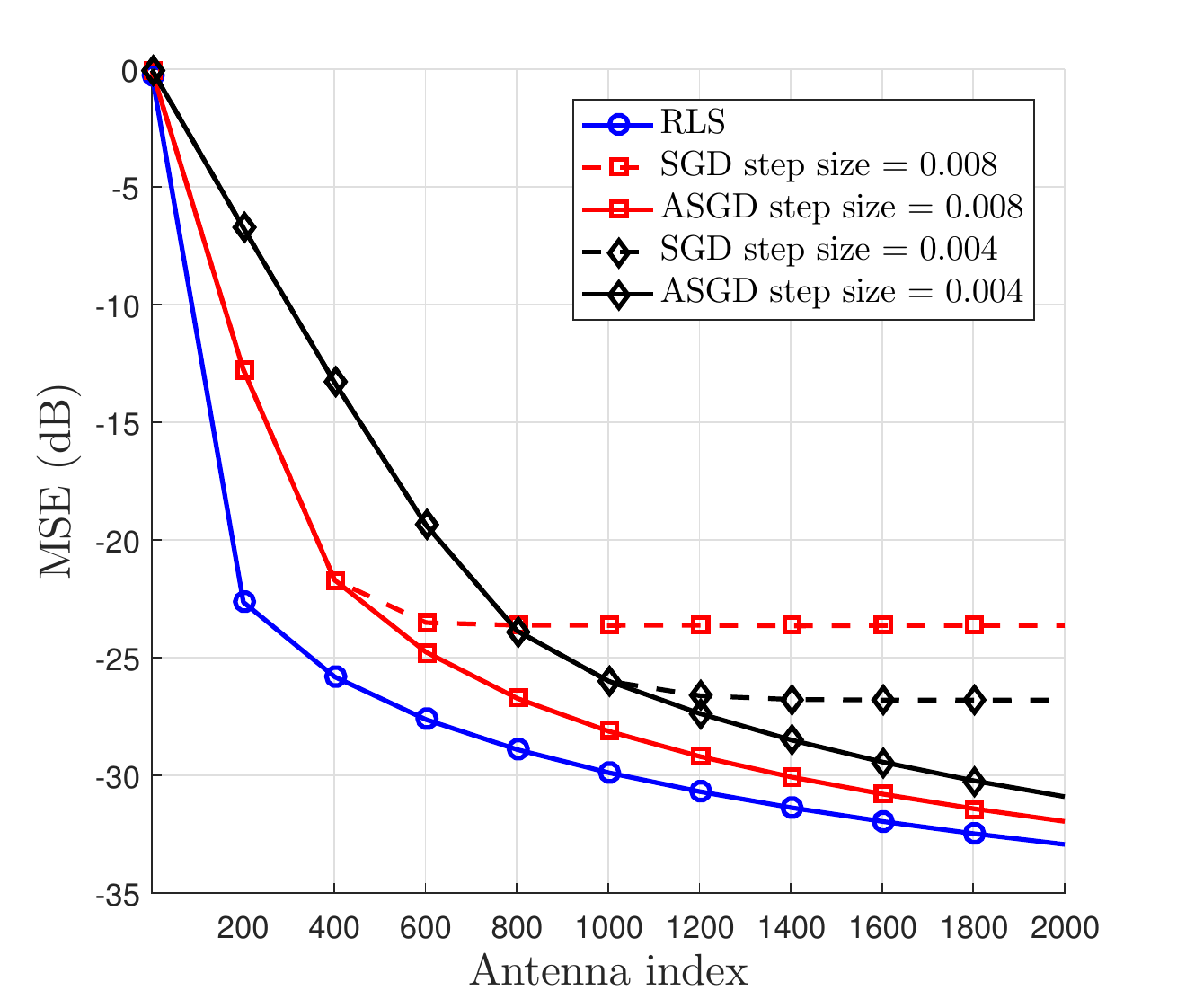}
		\label{fig:MSE_vs_antenna_RLS_SGD_ASGD_M2048}
	}
	\caption{MSE vs antenna index for RLS, SGD and ASGD for different step-size values. Left: M=256. $n_0$=150 and 75 for $\mu$=0.02 and 0.04 respectively.
		Right: M=2048. $n_0$=1000 and 400 for $\mu$=0.004 and 0.008 respectively. K=16 and SNR=12dB in all cases.}
	\vspace*{-3mm}
	\label{fig:MSE_vs_antenna_RLS_SGD_ASGD}
\end{figure*}

\begin{figure*}[t!]
	\centering
	\subfloat{
		\includegraphics[width=0.4\textwidth]{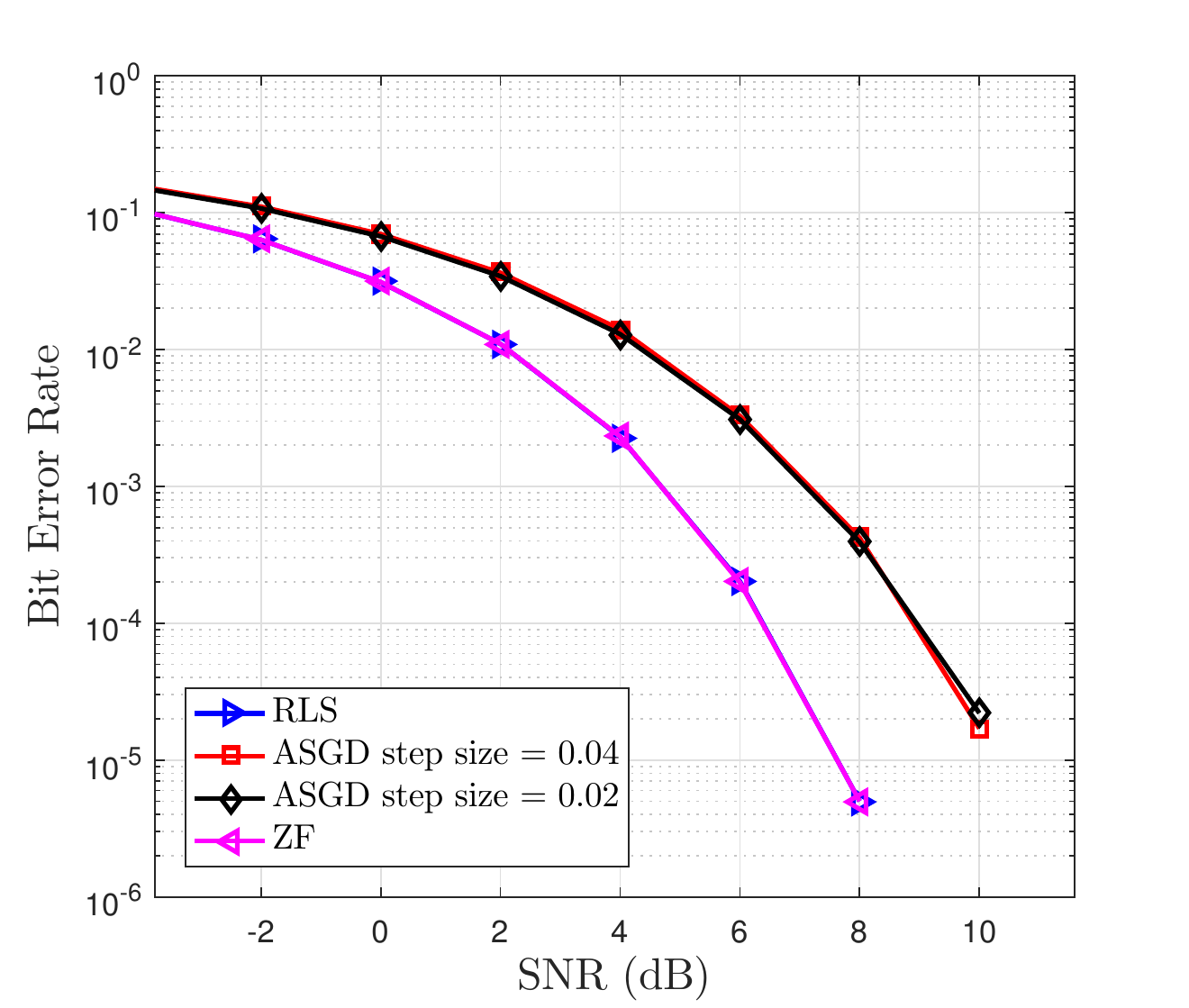}
		\label{fig:BER_vs_SNR_RLS_SGD_ASGD_M256_K16_16QAM}
	}
	\subfloat{
		\includegraphics[width=0.4\textwidth]{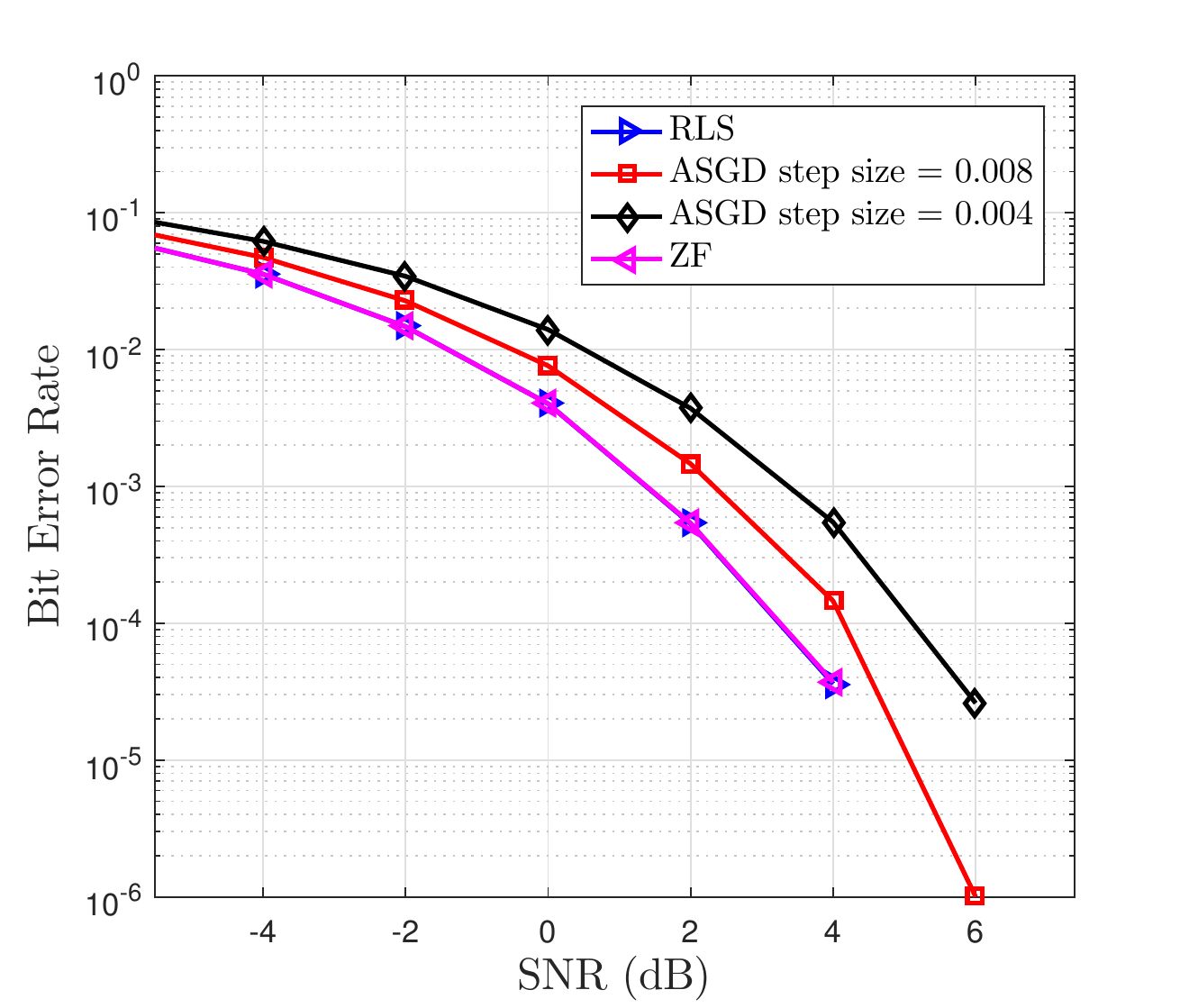}
		\label{fig:BER_vs_SNR_RLS_SGD_ASGD_M2048_K16_64QAM}
	}
	\caption{BER vs SNR for RLS, ASGD and ZF. Left: M=256, 16QAM. Right: M=2048, 64QAM. K=16 and SNR=12dB for both cases.}
	\vspace*{-3mm}
	\label{fig:BER_vs_SNR_RLS_SGD_ASGD}
\end{figure*}

We now turn our attention towards Figures \ref{fig:MSE_vs_antenna_RLS_SGD_ASGD} and \ref{fig:BER_vs_SNR_RLS_SGD_ASGD} which compare RLS, SGD, and ASGD. When the SGD sequence is averaged, MSE and BER curves get closer for different step-sizes, making the algorithm robust against non-optimal step-size selection. The selection of $n_0$ also has an impact, but less compared to non-optimal step-size in SGD.
As shown in Figure \ref{fig:BER_vs_SNR_RLS_SGD_ASGD}, RLS meets ZF \eqref{eq:LS_sol} performance, as it is optimal for a Gaussian noise source \cite{polyak}. For large $M$, performance of RLS and ASGD converge due to ASGD's asymptotically optimal rate property.
\subsection{Strengths of Daisy-Chain Topology}
\begin{figure*}\centering
	\includegraphics[width=0.79\linewidth]{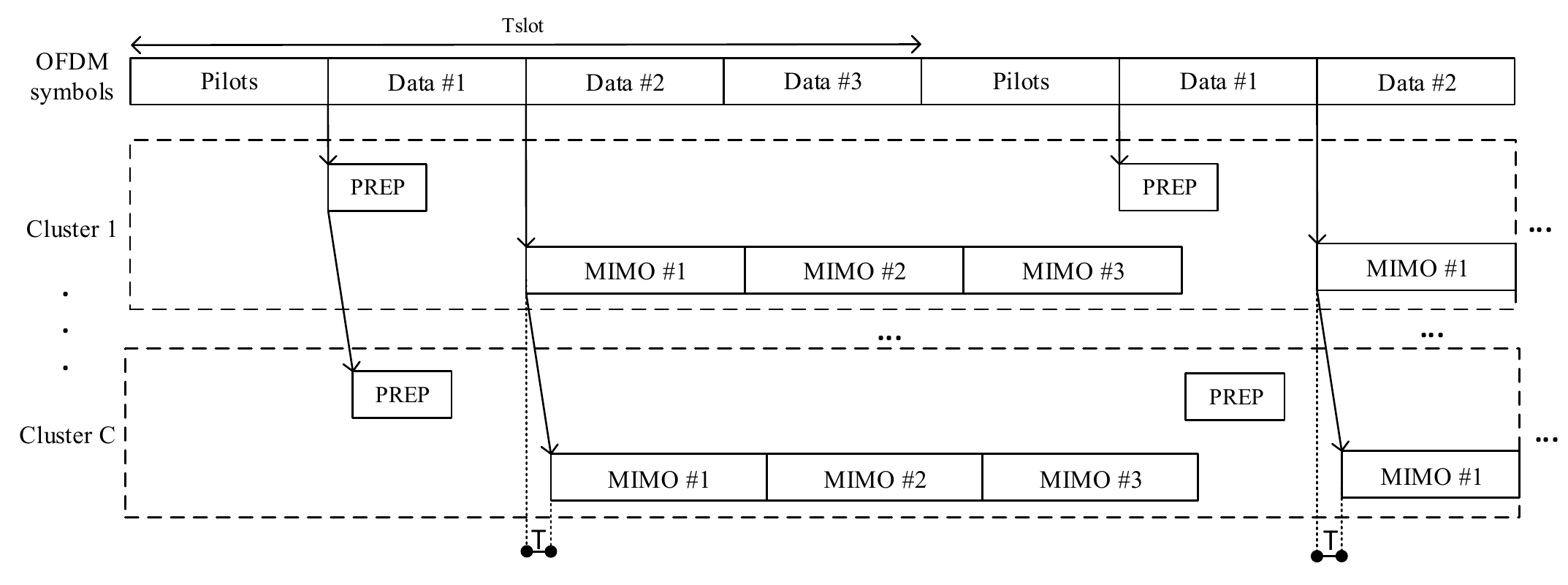}
	\caption{Time diagram representing cluster activities during one uplink slot with 4 OFDM symbol: one pilot and three data symbols. Only cluster 1 and C (last one) are represented for simplicity. Two types of activities are shown per cluster. The first one represents Pre-processing stage (PREP), only if RLS is used. The second one is the MIMO activity, which involves detection. First cluster start processing first RE after complete reception of Data 1. Once such RE is processed, it is then passed to next cluster for further processing and this is repeated successively through all clusters. As it is shown in the figure, there is a delay (T) for the starting time in cluster C compared to first cluster. T needs to be small enough to meet latency constraints.}
	\label{fig:time_diagram}
\end{figure*}
Ostensibly, it may come across as if our daisy-chain solution incurs a latency penalty. This is, however, not the case as the detection process over time and/or frequency can be pipelined. While cluster 2 is processing data at subcarrier, say, $f_0$, cluster 1 can process data at subcarrier $f_0+1$. In the next iteration, cluster 2 processes data at subcarrier $f_0+1$, etc. See Figure \ref{fig:time_diagram} for a graphical visualization of the pipelining procedure.

Further, our daisy-chain solution allows for a power save since if a cluster $n$ regards its incoming estimate to be sufficiently good, then it can do one out of at least two things, 1) set $\hat{\mathbf{s}}_{n+1}=\hat{\mathbf{s}}_n$, or 2) send the incoming estimate $\hat{\mathbf{s}}_n$ to the baseband processing node, thereby terminating the detection procedure. The former has the advantage over the latter that only the last cluster needs to be connected to the baseband processing unit. Further, an indication whether or not the incoming estimate is of sufficiently good quality can be obtained, e.g. for RLS, by the value $\varepsilon_n$ in \eqref{eq:RLS_sn}.

Finally, our topology is flexible so that additional antenna clusters can be added in a plug-and-play fashion. For example, in order to double the number of antennas, it is merely required to disconnect the cable between the last cluster and the baseband processing unit, connect that very cable to the last cluster of the added antenna array, and connect the two arrays. This will solely impact software scheduling at the baseband processing unit, but not the hardware as would have been the case for the centralized topology in Figure \ref{fig:BS_centralized}.

\subsection{Interconnection Data-Rate}
\label{section:interconnection}
In order to estimate the expected data-rate in the proposed architecture, we can assume an OFDM-based frame structure based on slots. Each slot is made by $N_{\mathrm{slot}}$ consecutive OFDM symbols with duration $T_{\mathrm{ofdm}}$. Each symbol contains $N_{\mathrm{u}}$ subcarriers (an RE in OFDM) to carry user data. We can determine the average input/output data rate in the uplink for each of the clusters during a certain slot for SGD as follows
\begin{equation}
\bar{R}_\mathrm{SGD} = \frac{K \cdot w_{\mathrm{s}} \cdot N_{\mathrm{u}} \cdot N_{\mathrm{UL}}}{T_{\mathrm{slot}}} = \alpha \cdot \frac{K \cdot w_{\mathrm{s}} \cdot N_{\mathrm{u}}}{T_{\mathrm{ofdm}}},
\label{eq:R_SGD}
\end{equation}
where $T_{\mathrm{slot}}$ is the slot duration, $N_{\mathrm{UL}}$ is the number of OFDM symbols allocated for UL data in a slot, $w_{\mathrm{s}}$ is the number of bits used to represent each element in the sequence of estimates ($\hat{\mathbf{s}}_n$) and $\alpha = \frac{N_{\mathrm{UL}}}{T_{\mathrm{slot}}}$ represents the fraction of time spent in UL within the slot, so $0 \leq \alpha \leq 1$. In Figure \ref{fig:BS_decentralized}, $\bar{R}_\mathrm{SGD}$ corresponds to $\textrm{R}$.

This analysis does not take into account the total amount of data that is generated (which depends on $M$) and needs to move through the structure, but only the data that moves between clusters (which depends on $K$) because it is the one that imposes physical constraints in the inter-cluster connections and may limit the scalability.

For ASGD, the averaged data rate is expected to be twice the one in SGD, because for each sequence element, two previous elements, $\hat{\mathbf{x}}_{n}$ and $\hat{\mathbf{s}}_{n-1}$, are needed as  can be observed in \eqref{eq:ASGD2}, and therefore
\begin{equation}
\bar{R}_\mathrm{ASGD} = 2 \cdot \bar{R}_\mathrm{SGD}.
\label{eq:R_ASGD}
\end{equation}

For RLS, the data-rate has two components, one due to the preprocessing stage and the other one due to each RE. During the first stage, matrix  $\mathbf{\Gamma}$ is passed from cluster to cluster. During the RE processing stage, data rate is the same as in SGD. The averaged data rate for RLS is calculated as
\begin{equation}
\begin{aligned}
\bar{R}_\mathrm{RLS} &= \frac{N_\mathrm{CB} \cdot K^2 \cdot w_{\gamma}}{T_\mathrm{slot}} + \frac{K \cdot w_\mathrm{s} \cdot N_\mathrm{u} \cdot N_\mathrm{UL}}{T_\mathrm{slot}} \\
&= \frac{N_\mathrm{u} \cdot N_\mathrm{slot}}{S_\mathrm{CB}} \cdot \frac{K^2 \cdot w_{\gamma}}{T_\mathrm{ofdm} \cdot N_\mathrm{slot}} + \alpha \cdot \frac{K \cdot w_\mathrm{s} \cdot N_\mathrm{u}}{T_\mathrm{ofdm}} \\
&= \alpha \cdot \frac{K \cdot w_\mathrm{s} \cdot N_\mathrm{u}}{T_\mathrm{ofdm}} \cdot \left( 1 + \frac{\beta}{\alpha} \cdot \frac{K}{S_\mathrm{CB}} \right), \\
\end{aligned}
\label{eq:R_RLS}
\end{equation}
where $N_\mathrm{CB}$ is the number of CBs per slot, $S_\mathrm{CB}$ the number of REs in each CB, $w_{\gamma}$ the number of bits to represent each element in $\mathbf{\Gamma}$ and $\beta = \frac{w_{\gamma}}{w_\mathrm{s}}$. From \eqref{eq:R_RLS} it can be seen that $\bar{R}_\mathrm{RLS} > \bar{R}_\mathrm{SGD}$.

We can compare our proposed solution with another cluster-based decentralized architecture, but which relies on a central node to collect partial results, performing a low complexity operation, such as averaging, and broadcasting back the result to the clusters according to an iterative algorithm. This star topology has been proposed in \cite{cavallaro}. In this case, the central node will have $C$ bi-directional links with an average aggregated data rate per direction of
\begin{equation}
\bar{R}_\mathrm{star} = C \cdot n_\mathrm{iter} \cdot \bar{R}_\mathrm{SGD},
\label{eq:R_star_central}
\end{equation}
where $n_\mathrm{iter}$ is the number of iterations for the selected detection algorithm.
From \eqref{eq:R_star_central} we can observe that typically $\bar{R}_\mathrm{star} \gg \bar{R}_\mathrm{SGD}$.

In case of a fully centralized architecture as the one in \cite{lumami}, the interconnection data-rate depends linearly on $M$ as follows
\begin{equation}
\bar{R}_\mathrm{central} = \frac{M \cdot N_\mathrm{u} \cdot N_\mathrm{UL} \cdot w_\mathrm{sc}}{T_\mathrm{slot}} = \alpha \cdot \frac{M \cdot N_\mathrm{u} \cdot w_\mathrm{sc}}{T_\mathrm{ofdm}},
\label{eq:R_central}
\end{equation}
where $w_\mathrm{sc}$ is the number of bits representing a sample of the received signal $\mathbf{y}$. It is seen that \eqref{eq:R_central} cannot scale easily. $\bar{R}_\mathrm{central}$ corresponds to $\textrm{R}$ in Figure \ref{fig:BS_centralized}. Going from \eqref{eq:R_central} to \eqref{eq:R_SGD}, roughly reduces the data-rate by a factor $\frac{M}{K}$ (typically $\geq 10$ in Massive MIMO).

Table \ref{tab:system_metrics} shows date-rates for four scenarios. We assume the following parameters: $T_\mathrm{slot}=500 \mu s$, $w_\mathrm{s}=16$, $w_\mathrm{sc}=24$, $N_\mathrm{u} = 1200$, $N_\mathrm{slot}=7$, $N_\mathrm{UL}=6$, $\alpha = 6/7$, $\beta = 3/2$, $S_\mathrm{CB}=400$ and $n_\mathrm{iter}=3$. We can observe that the analyzed topology and algorithms achieve significantly lower interconnection data-rate than other architectures \cite{lumami}\cite{cavallaro}, enabling future scalability. As observed, for very-large arrays, RLS and ASGD require similar data-rates and have similar performance, but RLS requires a pre-processing stage and matrix manipulation that ASGD does not.\\
\begin{table}
	\renewcommand{\arraystretch}{1.3} 
	\caption{Data Rate comparison for different topologies / algorithms}
	\label{tab:system_metrics}
	\centering
	\begin{tabular}{l*{5}{c}}
		\hline
		$M$    & 128   & 256   & 512   & 1024\\
		$K$	   & 16    & 32    & 64    & 128 \\
		$C$	   & 8     & 8     & 16    & 16 \\
		$B$	   & 16    & 32    & 32    & 64 \\
		\hline
		$\bar{R}_\mathrm{SGD}$  & 439MB/s & 879MB/s & 1.7GB/s & 3.4GB/s\\
		$\bar{R}_\mathrm{RLS}$  & 470MB/s & 1.0GB/s & 2.2GB/s & 5.3GB/s\\
		$\bar{R}_\mathrm{ASGD}$ & 879MB/s & 1.7GB/s & 3.4GB/s & 6.8GB/s\\
		\hline
		$\bar{R}_\mathrm{star}$\cite{cavallaro} & 10.3GB/s & 10.3GB/s & 20.6GB/s & 20.6GB/s\\
		$\bar{R}_\mathrm{central}$\cite{lumami} & 5.1GB/s & 10.2GB/s & 20.4GB/s & 40.8GB/s\\
	\end{tabular}
\end{table}
\section{Conclusions}
\label{section:conclusions}
In this article we have introduced a base station uplink architecture for Massive MIMO and analyzed the main implementation bottleneck, the interconnection data-rate. We have proposed three algorithms and a fully decentralized topology for uplink detection, which alleviate this limitation. One of the algorithms (RLS) achieves approximate zero-forcing performance, while another (ASGD) is an approximation which converges to the former one for very large arrays. All of them are of low-complexity and do not require matrix inversion. An estimate of data-rate is also presented and compared with other architectures for different array-sizes and configurations, showing the benefits of the proposed solution.
\section*{Acknowledgment}
This work was supported by ELLIIT, the Excellence Center at Link\"{o}ping-Lund in Information Technlology.
\nocite {polyak2, book:Ljung,book:Kushner}
\bibliographystyle{IEEEtran}
\bibliography{IEEEabrv,sips2018}

\begin{thebibliography}{10}
\providecommand{\url}[1]{#1}
\csname url@samestyle\endcsname
\providecommand{\newblock}{\relax}
\providecommand{\bibinfo}[2]{#2}
\providecommand{\BIBentrySTDinterwordspacing}{\spaceskip=0pt\relax}
\providecommand{\BIBentryALTinterwordstretchfactor}{4}
\providecommand{\BIBentryALTinterwordspacing}{\spaceskip=\fontdimen2\font plus
\BIBentryALTinterwordstretchfactor\fontdimen3\font minus
  \fontdimen4\font\relax}
\providecommand{\BIBforeignlanguage}[2]{{%
\expandafter\ifx\csname l@#1\endcsname\relax
\typeout{** WARNING: IEEEtran.bst: No hyphenation pattern has been}%
\typeout{** loaded for the language `#1'. Using the pattern for}%
\typeout{** the default language instead.}%
\else
\language=\csname l@#1\endcsname
\fi
#2}}
\providecommand{\BIBdecl}{\relax}
\BIBdecl

\bibitem{marzetta}
T.~L. Marzetta, ``Noncooperative cellular wireless with unlimited numbers of
  base station antennas,'' \emph{IEEE Transactions on Wireless Communications},
  vol.~9, no.~11, pp. 3590--3600, November 2010.

\bibitem{argos}
\BIBentryALTinterwordspacing
C.~Shepard \emph{et~al.}, ``Argos: Practical many-antenna base stations,'' in
  \emph{Proceedings of the 18th Annual International Conference on Mobile
  Computing and Networking (Mobicom)}, New York, NY, USA, 2012, pp. 53--64.
  [Online]. Available: \url{http://doi.acm.org/10.1145/2348543.2348553}
\BIBentrySTDinterwordspacing

\bibitem{larsson}
E.~Bertilsson, O.~Gustafsson, and E.~G. Larsson, ``A scalable architecture for
  massive {MIMO} base stations using distributed processing,'' in \emph{2016
  50th Asilomar Conference on Signals, Systems and Computers}, Nov 2016, pp.
  864--868.

\bibitem{puglielli}
A.~Puglielli \emph{et~al.}, ``Design of energy- and cost-efficient massive
  {MIMO} arrays,'' \emph{Proceedings of the IEEE}, vol. 104, no.~3, pp.
  586--606, March 2016.

\bibitem{lumami}
S.~Malkowsky \emph{et~al.}, ``The world's first real-time testbed for massive
  {MIMO}: Design, implementation, and validation,'' \emph{IEEE Access}, vol.~5,
  pp. 9073--9088, 2017.

\bibitem{cavallaro}
K.~Li, R.~R. Sharan, Y.~Chen, T.~Goldstein, J.~R. Cavallaro, and C.~Studer,
  ``Decentralized baseband processing for massive {MU-MIMO} systems,''
  \emph{IEEE Journal on Emerging and Selected Topics in Circuits and Systems},
  vol.~7, no.~4, pp. 491--507, Dec 2017.

\bibitem{book:Ljung}
L.~Ljung and T.~S\"oderstr\"om, \emph{Theory and Practice of Recursive
  Identification}.\hskip 1em plus 0.5em minus 0.4em\relax The MIT Press, 1983.

\bibitem{book:Kushner}
H.~J. Kushner and G.~G. Yin, \emph{Stochastic Approximation and Recursive
  Algorithms and Applications}, 2nd~ed.\hskip 1em plus 0.5em minus 0.4em\relax
  Springer, 2003.

\bibitem{polyak}
B.~Polyak and A.~B.~Juditsky, ``Acceleration of stochastic approximation by
  averaging,'' \emph{SIAM Journal on Control and Optimization}, vol.~30, pp.
  838--855, July 1992.

\bibitem{polyak2}
B.~Polyak and Y.~Z.~Tsypkin, ``Adaptive estimation algorithms (convergence,
  optimality, stability),'' \emph{Automation and Remote Control}, vol.~40, pp.
  378--390, March 1979.

\end{thebibliography}
\end{document}